\documentclass[useAMS,usenatbib]{mn2e}

\usepackage{graphicx}
\usepackage{txfonts}

\title[Collision of a Mars-sized planet to form the Moon]{On the probability of the collision of a Mars-sized planet with
the Earth to form the Moon}
\author[R.\ Dvorak]{Rudolf Dvorak\thanks{E-mail:
rudolf.dvorak@univie.ac.at}, Birgit Loibnegger, and Thomas I.\ Maindl\\
Department of Astrophysics, University of Vienna, T\"urkenschanzstrasse 17, 1180 Vienna\\
}

\begin{document}

\date{}
\pagerange{\pageref{firstpage}--\pageref{lastpage}} \pubyear{2015}

\maketitle

\label{firstpage}

\begin{abstract}
The problem  of the formation of the Moon is still not explained satisfactorily.
While it is a generally accepted scenario that the last giant impact on Earth between some 50  to 100 million years 
after the
starting of the formation of the terrestrial planets formed our natural 
satellite, there are still
many open questions like the isotopic composition which is identical for
these two bodies. In our investigation we will not deal with these problems of
chemical composition but rather undertake a purely dynamical study to find out
the probability of a Mars-sized body to collide with the Earth shortly after
the formation of the Earth-like planets. For that we assume an additional
massive body between Venus and Earth, respectively Earth and Mars which formed
there at the same time as the other terrestrial planets. We have undertaken 
massive n-body integrations of such a planetary system with 4 inner
planets (we excluded Mercury but assumed one additional body as mentioned 
before) for up to tens of millions of years. Our results led to a statistical
estimation of the collision velocities as well as the collision angles 
which will then serve as the basis of further investigation with 
detailed SPH computations. We find a most probable origin of the Earth impactor at a semi-major axis of approx.\ 1.16\,AU.
 
\end{abstract}


\section {Introduction}
An  assumed giant impact of an additional Mars-sized object (Theia) onto the Earth 
could have led to the formation of the
Moon after the planets already had their actual mass and no more gas was
present in the Solar System. Many recent publications deal with this topic,
e.g., \cite{Asp14}, \cite{Nak15}, \cite{Qua14}, \cite{Jac14a}, \cite{Jac14b} [=JM],
\cite{Izi14}, since the first ideas developed by
\cite{Har75}, \cite{Cam76}, and later \cite{Can01}.
Detailed collision scenaria were studied e.g., by \cite{Cam97},
\cite{Can04}, \cite{Can08}, and 
\cite{Can13} where the
collision was modelled with the aid of sophisticated SPH codes. In a most
recent article \cite{Kai15} [=KC] the authors concentrate on the feeding zone
of the planet to form with respect to the planet's volatile inventory and
isotopic composition. Because of the highly random outcome they ask the
question of how deterministic the outcome of the planetary formation is. In
fact the correspondence of the results of the different model computations of
n-body codes is very small. Most of these modelisations have been
undertaken to understand the architecture of our Solar System, which results
only from a subset of the chosen initial conditions.
In KC the authors estimated the likelihood that the mass of Theia could
be equal to the mass of the Earth, but the probability is rather low. Their
results coincide well with JM who estimated the collision probability of bodies
with comparable masses as being low. According to these results we have fixed
the mass of the additional planet (the `projectile planet') to 
$m_\mathrm{Mars}\/$ for our computations. Other investigations aimed for high velocity
encounters e.g. the one by \cite{Cuk12} who assumed 
high velocity collisions for smaller masses of Theia ($0.025\, m_\mathrm{Earth} <
m_\mathrm{Theia} < 0.05\, m_\mathrm{Earth}$), but the results of KC show that such an event may not be
very probable because a spin rate of the Earth of the order of 2 hours can
only be achieved by big impactors -- and these events are rare. Because of the
same reason the scenario proposed by \cite{Reu12} where they look for a
steeper collision angle is not very probable. KC undertake 150 different
simulations with 3 different underlying models: a first model with Jupiter
and Saturn on circular orbits, a second one with initially small eccentricities
of Jupiter and Saturn, and a third one according to the model of \cite{Han09}. Whereas in the first two models 100 self interacting bodies
(distributed between 0.5 and 4 AU with small eccentricities) were integrated
which then end  up as planets, the last one starts with 400 embryos in an
annulus between 0.7 and 1 AU and -- according to the authors -- represent more or
less the outcome of the Grand Tack model \citep{Wal12}. It is therefore
appropriate to make such a study -- which is orientated versus the collision of a 
Mars-sized object with the Earth -- on the basis of these results.

\subsection{A possible formation of a `Theia' in the early planetary system}

\begin{figure}
\begin{center}
\includegraphics[width=5.0cm,angle=270]{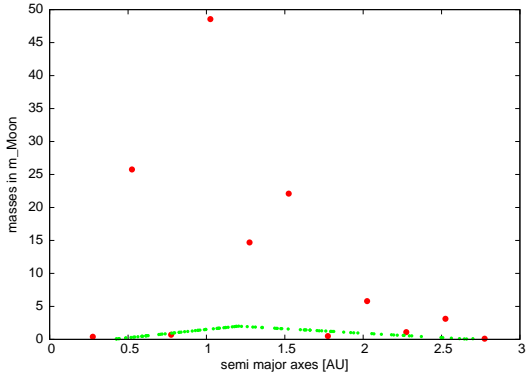}
\caption{Semi-major axes versus the mass of formed planets in an example where
  the initial conditions  where chosen after the Grand Tack scenario \citep{Wal12}.}
\label{ss-ex}
\end{center}
\end{figure}

\begin{figure}
\begin{center}
\includegraphics[width=5.0cm,angle=270]{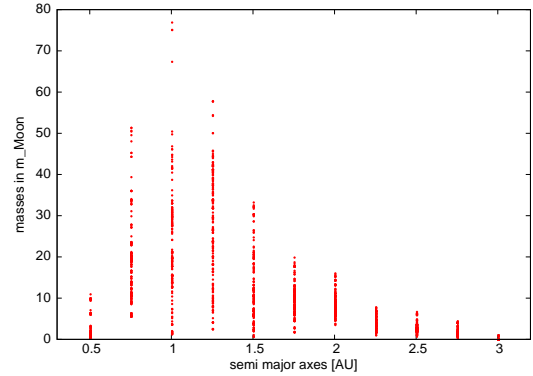}
\caption{Combined results of 16 different planet forming simulations; axes
  like in Fig.~\ref{ss-ex}.}
\label{ss-all}
\end{center}
\end{figure}

It is well known that the outcome of computations of the development of the
early Solar System depends highly on the initial conditions after the gas in
the disk disappeared. Several different approaches lead to different
`planetary systems' although all these attempts have been undertaken to 
understand the architecture of our system (e.g. \citep{Han09,Izi14}). Our assumptions is based on the outcome of the 
Grand Tack \citep{Wal12} where after the inward migration of 
Jupiter and Saturn a later outward migration triggered the
formation of the terrestrial planets. In our numerical integrations we started
with these two  gas giants in their actual position and 100
planetesimals randomly distributed 
between $0.4\,\mathrm{AU} < a_\mathrm{planetesimal} < 2.7\,\mathrm{AU}$ with masses in the order of the
Moon. In Fig.~\ref{ss-ex} we plotted the results of one out of 16 simulations 
where the architecture of this simulated planetary system turned out to be
close to the one of our Solar System. Nevertheless there are two important
differences: an additional planet
(about the size of Mars) between the `Earth' (here with only 60\,\% of its
actual mass) and a `Mars' (with more than the double of its actual mass); the
planet at 0.5\,AU can be seen as a Venus equivalent. 
While this is just one example of resulting configurations we use it to motivate 
our choice of initial conditions described in the section below:
a Mars-sized planet between Earth and Mars. The example chosen is not fully 
artificial when we look at Fig.~\ref{ss-all}, where all the simulations are 
combined into one graph which shows the accumulation of planets around 1\,AU.

\section{Dynamical model and numerical setup}
\label{sect:setup}

Our dynamical models for the early Solar System where chosen in the following
way:

\begin{itemize}
\item {\bf Model 1 [M1],} consisting only of the terrestrial planets excluding Mercury and an additional planet in between the orbits of Earth and Venus.
\item {\bf Model 2 [M2],} consisting of the terrestrial planets excluding Mercury, the gas giants Jupiter and Saturn, and an additional planet in between the orbits of Earth and Mars.
\end{itemize}

\noindent Because our tests have shown that for the additional planet between Venus
and Earth the influence of Jupiter and Saturn can be neglected we just used a 
5-body problem for {\bf M1}.
For both models we adopted as initial conditions the current osculating 
elements of the planets and added a Mars-sized object, {\bf Theia}, with semi-major axes $0.8\,\mathrm{AU} < a_\mathrm{Theia} < 0.94\,\mathrm{AU}$ ({\bf M1}) and $1.06\,\mathrm{AU} < a_\mathrm{Theia} < 1.4\,\mathrm{AU}$ ({\bf M2}), respectively. Between scenarios we varied Theia's semi-major axes with $\delta a_\mathrm{Theia} = 0.005\,\mathrm{AU}$.
The reason of taking the orbital elements
of the planets as they are today is the following: we know
that for the billions of years into the past the orbits were the same (e.g. 
\cite{Las96}) or only slightly different. The additional planet -- the possible 
impactor -- should then be in a quasi stable orbit after the formation of 
the planetary system. To find such an orbit we did not vary the
eccentricity nor the inclination and set them to $e_\mathrm{ini}=0.075$  and 
$i=2^{\circ}$.  For the other orbital elements of Theia we
used randomly chosen  values between  $0^\circ$ and $360^{\circ}$. 
For every fixed semi-major axis 25 such osculating elements were
computed as initial conditions. For achieving the highest possible 
precision with respect to the collision angle and velocity the integration 
method was the one we have used for many years for most of our computations
\citep{Dvo03,Dvo15,Gal13}. The Lie-integration has an automatic step-size control and is well 
adapted for such kind of computations \citep{Han84,Egg10,Lic84,Del85}.
With regard to the formation process of the terrestrial
planets and the estimated time of collision of Theia with the Earth of $95 \pm
32$\,Myr (\cite{Jac14b}) the integration time was up to 50 Myrs. 

\section{Evolution of two selected samples}

To demonstrate the sensitivity of the dynamical evolution in this regions  
we show a stable seeming orbit and another one very rapidly becoming
unstable. Both had initial conditions outside the Earth orbit with only slightly
different semi-major axis, and the same 
initial conditions for the other orbital elements.
 
\begin{figure}
\begin{center}
\includegraphics[width=5.0cm,angle=270]{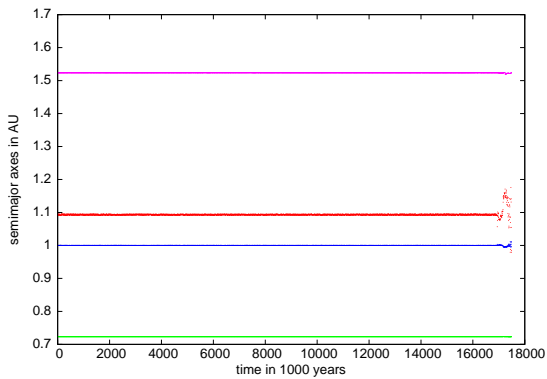}
\caption{Time development of the semi-major axes of the terrestrial planets and additional planet Theia leading to a collision with the Earth.}
\label{stable}
\end{center}
\end{figure}

\begin{figure}
\begin{center}
\includegraphics[width=5.0cm,angle=270]{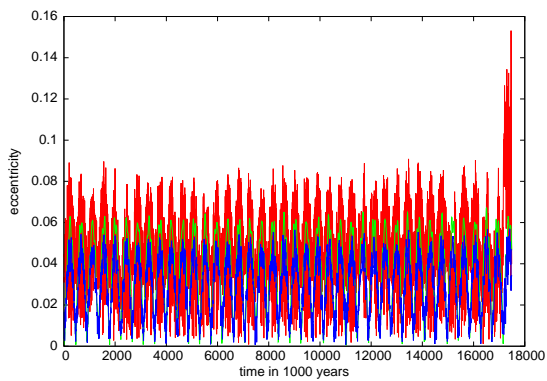}
\caption{Time development of the eccentricities of the Earth, Venus, and 
	in addition the one of 
Theia (in red) from the example in Fig.~\ref{stable}. Please see text for more.}
\label{60e}
\end{center}
\end{figure}

\begin{figure}
\begin{center}
\includegraphics[width=5.0cm,angle=270]{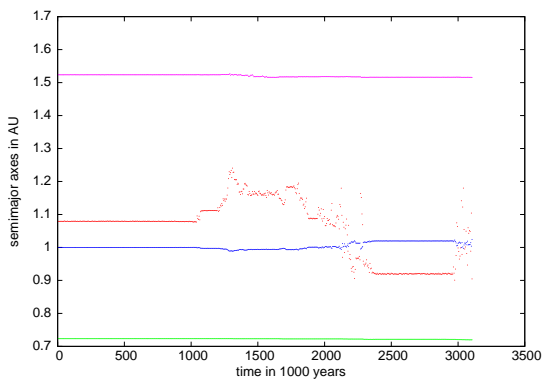}
\caption{Time development of the semi-major axes of a highly chaotic 
orbit leading to an escape after 3\,Myrs.}
\label{chaos}
\end{center}
\end{figure}

\begin{figure}
\begin{center}
\includegraphics[width=5.0cm,angle=270]{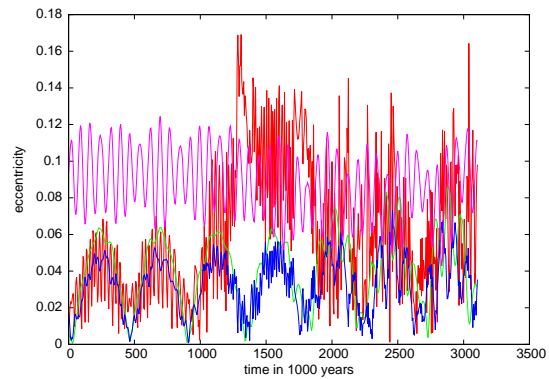}
\caption{Time development of the eccentricity (y-axis) of a highly chaotic orbit of
  Theia from Fig.~\ref{chaos} with a subsequent
  escape after a close encounter with the Earth; the eccentricity of Mars is
the regular curve close to $e=0.1\/$}
\label{23e}
\end{center}
\end{figure}

In Fig.~\ref{stable} and  Fig.~\ref{chaos}  we show two examples of 
the development of the semi-major axes of a fictitious Theia and the 
terrestrial planets, one for a stable orbit  and another
one which is chaotic very soon. The very regular variation of the
eccentricity of this stable orbit of Theia is depicted in Fig.~\ref{60e} ($0 < e_\mathrm{Theia} < 0.08$) 
which ends in a sudden increase up to $e=0.15$, a close approach and even a
collision with the Earth after 17\,Myrs.
In the other example (Fig.~\ref{60e}) we show a chaotic orbit suffering from multiple close
encounters with the Earth visible through the chaotic signal of Theia's semi-major
axis already after 1\,Myr. Caused by a sequence of very close encounters between Earth and Theia
the orbit of the fictitious planet jumps inside the orbit of the Earth after 2.2\,Myrs.
Also the Earth is suffering from these repeatedly close encounters and moves a
little outside which can be explained by the conservation of the momentum of
this planet pair. After a  capture close to the 6:5 MMR between them (2.2\,Mrys) a final
collision with the Earth ends the lifetime of Theia. Mars
and Venus are not affected at all in this example.

\section{Impact parameters}

\begin{figure}
\begin{center}
\includegraphics[width=5cm]{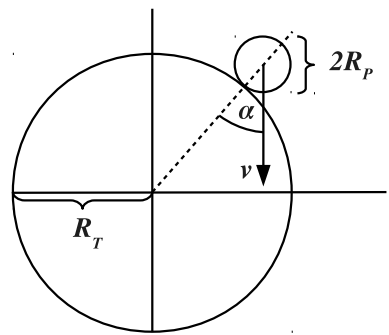}
\caption{Schematic view of the collision parameters impact angle and  impact  velocity for an  Earth-colliding small planet.}
\label{param}
\end{center}
\end{figure}

The goal of these investigations is to determine the collision 
probability for an additional Mars-sized planet Theia which could have formed
during the early formation (e.g. Grand Tack scenario) between the Earth and 
Mars or between Earth and Venus. In addition to this we study the impact
velocity as well as the impact angle (Fig.~\ref{param}) -- expressed by the impact parameter in
different studies \citep{Lei12,Mai14}.

\begin{figure}
\begin{center}
\includegraphics[width=7.5cm]{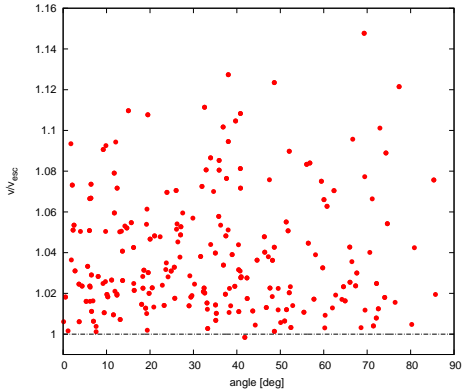}
\caption{Collision parameters impact angle versus impact velocity for the
  Earth-colliding Theias.}
\label{angles}
\end{center}
\end{figure}

Together
with the mass and composition of the two colliding bodies these two quantities determine the outcome of such
a cosmic catastrophic event. In
Fig.~\ref{angles} we show the different impact parameters according to our simulations, 
where one can see that most
of the encounter velocities are just above the escape 
velocity of the two planets ($v_\mathrm{esc} \sim 9\,\mathrm{km/s}$) and only one 
impact happens with a 15 percent higher velocity.
The distribution of the impact angles is asymmetric with slightly more head-on 
collisions; we can explain it by the fact that close to the Earth Theia 
suffers from a strong acceleration toward the center of the larger planet (we
remember the mass ratio $\mu = 0.1$).
  
The actual collision outcome in terms of fragmentation strongly depends on the impact velocities, impact angles, and the mass ratio \citep{Lei12,Mai14b}. The number of fragments varies with the impact angle such that even for relatively high impact velocities strongly inclined collisions may lead to two major survivors (hit-and-run) whereas head-on impacts may destroy the involved bodies. In our scenarios the low impact velocities suggest surviving bodies and probably a Moon forming from a debris ring.

\section{The overall picture}

\begin{figure}
\begin{center}
\includegraphics[width=7.5cm]{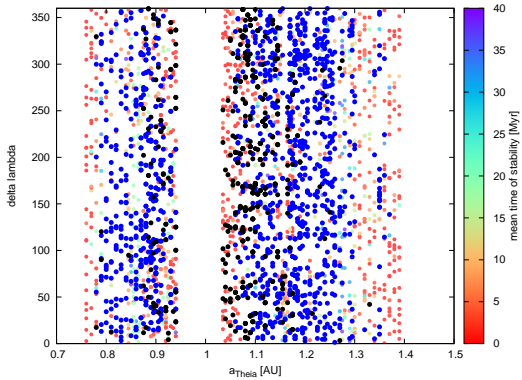}
\caption{Initial condition diagram: differences in mean logitudes with respect
  to the Earth versus the semi-major axis. The color points are coded with
  respect to their time of stability.}
\label{punk}
\end{center}
\end{figure}

For the integrated several thousands of orbits and the chosen grid of
semi-major axes a fixed eccentricity and a fixed value of the
inclination was chosen; the initial mean longitude was varied 
randomly (see Sect.~\ref{sect:setup}). In Fig.~\ref{punk} we show the combined 
results for all computed orbits by plotting the initial difference in mean 
longitudes ($\lambda_\mathrm{Theia}-\lambda_\mathrm{Earth}$) versus semi-major axes and color
coding the `stability character' of the orbit which was classified 
as follows: red to light blue circles according to the escape time. These
color points stand
for an unstable orbit either due a close encounter with Venus (inner part) or due
to a close encounter with Mars or even with the Earth. We note such 
an orbit as one typical example is depicted in Fig.~\ref{chaos}. 
Dark blue circles stand for stable orbits for the whole integration time; the
black dots stand for a collisions with the Earth.

\begin{figure}
\begin{center}
\includegraphics[width=7.5cm]{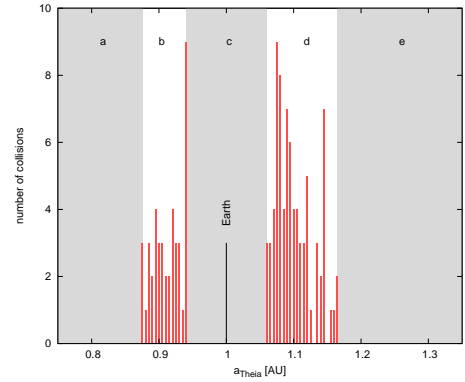}
\caption{Schematic view of the domain of the terrestrial planets with respect
  to the semi-major axis. In the regions 
  {\bf b} and {\bf d} we plotted the number of observed collisions with 
the Earth; for more see text.}
\label{abcde}
\end{center}
\end{figure}

According to the chosen initial conditions in semi-major axis we divided the
whole domain where the terrestrial planets move in 5 regions: 
{\bf (a)} $a_{Theia} \le 0.875\,\mathrm{AU}$, {\bf (b)} $0.875\,\mathrm{AU} <
a_{Theia} \le 0.94\,\mathrm{AU}$, {\bf (c)} the region around the Earth, {\bf
  (d)} $1.06\,\mathrm{AU} < a_{Theia} \le 1.165\,\mathrm{AU}$, and {\bf (e)} 
$a_{Theia} > 1.165\,\mathrm{AU}$. In regions {\bf b} and
{\bf d} we plotted the number of collisions with the Earth for each of the 25 initial mean longitudes per initial $a_\mathrm{Theia}\/$ (Fig.~\ref{abcde}). We observe that the maximum number of collisions with the Earth is about 36\,\% (9 out of 25 initial conditions), which occurs once for regions {\bf b} ($a_\mathrm{Theia}= 0.95\,\mathrm{AU}$) and {\bf d} ($a_\mathrm{Theia}= 1.075\,\mathrm{AU}$), respectively.

\begin{figure}
\begin{center}
\includegraphics[width=7.5cm]{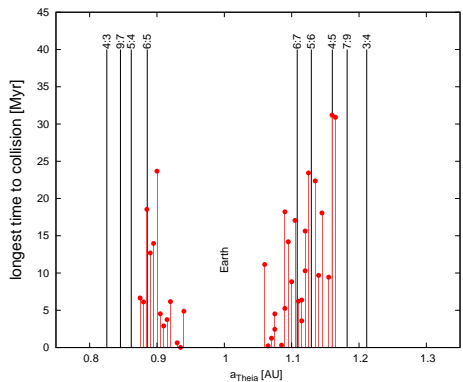}
\caption{Mean Motion Resonances of Theia with the Earth in regions {\bf b} and
  {\bf d}. Also the longest time before a collision versus semi-major axis is
  plotted here; for more see text.}
\label{mmr}
\end{center}
\end{figure}

In regions {\bf a} and 
{\bf e} many mean motion resonances (MMR) with the Earth act to destabilize 
the orbits of a hypothetical Theia, see Fig.~\ref{mmr} which also shows the most important MMRs together
with the longest duration of stability of the collision orbits in regions   
{\bf b} and {\bf d}. One can see that the number of
Earth collisions increases with the decreasing distance to the Earth. We did
not make computations close to the Earth (inside and outside its orbit)
-- region {\bf c} -- with the exceptions of some sample orbits, which have shown that
almost all are destabilized very soon after a close encounter or even a
collision. Because our interest was to find 
collisional orbits which were stable for several tens of millions of years in between
Earth and Venus and  also between Earth and Mars we do not show these results. It is easy
to understand that only then an additional planet may have formed in the early
formation stages of our planetary system; we showed even an example from our
own computations of the formation of the planets in Fig.\ref{ss-ex} where a
Theia like planet was formed outside the Earths orbit.

\begin{table}
\begin{center}
\begin{tabular}{|c|c|l|l|l|} \hline
\hline
\hline
$a_\mathrm{Theia}$&CT-mean&CT-max&CT-min&n \\
    0.875 &   3.41 &  6.648&1.555       &3\\
    0.880 &   6.13 &  6.129&6.129       &1\\
    0.885 &  12.87 & 18.542&2.892       &3\\
    0.890 &   8.52 & 12.691&4.342       &2\\
    0.895 &   7.02 & 13.973&3.332       &4\\
    0.900 &   8.21 & 23.669&0.313       &3\\
    0.905 &   1.83 &  4.528&0.114       &3\\
    0.910 &   2.06 &  2.904&1.216       &2\\
    0.915 &   1.92 &  3.747&0.090       &2\\
    0.920 &   4.22 &  6.168&1.628       &4\\
    0.925 &   9.76 & 15.685&0.057       &3\\
    0.930 &   0.26 &  0.626&0.051       &3\\
    0.935 &   0.02 &  0.024&0.024       &1\\
    0.940 &   0.97 &  4.872&0.002       &9\\
\hline
\end{tabular}
\end{center}
\caption{Collisions in region {\bf b}, see text for details.}
\label{allcoll1}
\end{table}

\begin{figure}
\begin{center}
\includegraphics[width=7.5cm]{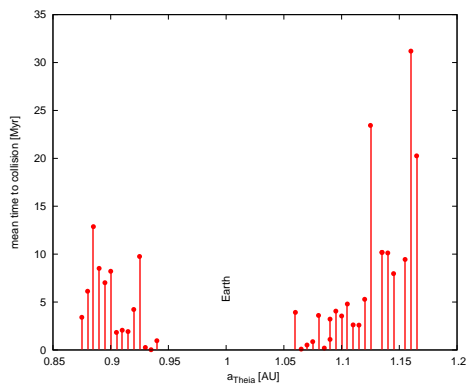}
\caption{Mean time to collisions in Myrs versus semi-major axis of Theia in AU.}
\label{meancoll}
\end{center}
\end{figure}

It is obvious that close to
the inner region {\bf c} the stability time is short
because the Earth is relatively close by, whereas we find times up to 30\,Myrs of
stability for an orbit before an escape in region {\bf d}.
The results for the mean time before a collision occurs are shown in Fig.~\ref{meancoll}.
It is obvious, that close to the inner region {\bf c} this time is short
because the Earth is relatively close by, whereas this mean collision time
increases in regions {\bf b} and {\bf d}.

\begin{table}
\begin{center}
\begin{tabular}{|c|c|l|l|l|} \hline
\hline
\hline
$a_\mathrm{Theia}$ & CT-mean & CT-max & CT-min & n \\
    1.065 &   0.17 &  0.235&0.013       &2\\
    1.070 &   0.52 &  1.255&0.008       &4\\
    1.075 &   0.86 &  4.518&0.010       &9\\
    1.080 &   3.60 & 10.471&0.017       &8\\
    1.085 &  11.10 &  0.314&0.104       &4\\
    1.090 &   3.22 & 18.218&0.023       &7\\
    1.095 &   8.21 & 14.190&0.205       &6\\
    1.100 &  15.20 &  8.838&0.117       &4\\
    1.105 &   2.30 & 17.066&0.426       &4\\
    1.110 &   2.62 &  6.225&0.392       &3\\
    1.115 &   2.59 &  3.580&1.259       &3\\
    1.120 &   5.28 & 15.63 &1.529       &5\\
    1.125 &  23.43 & 23.435&23.435      &1\\
    1.130 &   -    &-      &-           &-\\
    1.135 &  10.19 & 14.778&1.299       &3\\
    1.140 &  10.12 &  9.691&5.27        &2\\
    1.145 &   7.97 & 18.064&3.525       &7\\
    1.150 &      - & -     &-           &-\\
    1.155 &   9.44 &  9.443&9.443       &1\\
    1.160 &  31.19 & 31.193&31.193      &1\\
    1.165 &  20.26 & 30.895&9.617       &2\\
\hline
\end{tabular}
\end{center}
\caption{Collisions in region {\bf d}}
\label{allcoll2}
\end{table}

\section{Conclusions}

\begin{table}
\begin{center}
\begin{tabular}{|c|c|l|l|l|}
\hline
region    &  $a_\mathrm{Theia}$ in AU & stable  & eject & collision \\
a  &  0.750-0.875   &  50.50 &47.25& 2.25 \\ 
b  &  0.875-0.940 & 47.08  & 40.61   & 12.31 \\
c  &  0.940-1.060 &   -  &  - & -  \\
d  &  1.060-1.165 & 44.73   &  29.27  & 26.00\\
e  &  1.165-1.350 & 52.16   &  45.12  & 2.72 \\
\hline
\end{tabular}
\end{center}
\caption{Statistics of collisions (units percents)}
\label{allcoll3}
\end{table}

Thousands of orbits of a hypothetical additional planet Theia in
the early phase of the Solar System were integrated and classified
according their dynamics. The aim was to find an orbit stable for sufficiently
long time (tens of Myrs) inside or outside the orbit of the Earth which then
could lead to a collision building a companion of the Earth.
The impactor was assumed having the mass of Mars which then -- after a
collision -- could lead to an additional body like our Moon.  

Tabs.~\ref{allcoll1} and~\ref{allcoll2} summarize our results with respect to
the mean, maximum and minimum collision times for the inner region {\bf b} and
the outer region {\bf d}, respectively. The last columns show the number of 
Earth collisions of Theia, which is always 
smaller than 40 percent of the 25 integrated orbits for one fixed 
semi-major axis. 
Finally, Tab.~\ref{allcoll3} shows the overall statistics.
Approximately 50\,\% of all integrated orbits (in total about 2000, the
different tests in regions {\bf a} and {\bf e} included) turned out to be
stable up to the chosen integration time of up to 50 Myrs. The same almost 50\,\% in
these two regions escape due to close encounters either with Venus or Mars (we
did not count the number of Venus and Mars collisions) and
only 2\,\% suffered from impacts on Earth. In regions {\bf b} again  approx.\ 50\,\%
turned out to be stable, but the number of Earth colliders was increased to 12\,\%. Region {\bf d} is the best candidate for a collision of Theia with the
Earth after millions of years of stability: 26\,\% of such colliders were
observed here and especially around 1.16\,AU (compare Fig.~\ref{mmr}) the chance
is high that the Moon-producing planet Theia was formed here together with the
other terrestrial planets (compare Fig.~\ref{ss-ex}). 
We plan to make more of these numerical experiments on a finer grid and for
longer integration times and furthermore we will combine these results with detailed
computations of the collisions (SPH codes); all this will bring us a step further
in the knowledge of the formation of our Moon. 

\section{acknowledgements}

Special thanks for financial support are due to the FWF, project  S-11603-N16.

\end{document}